\documentclass[11pt,amstex,epsfig]{article}
\usepackage{amsmath,epsfig}

\font\affiliation=cmssi10

\begin{document} 
%\tighten 
\title{The complete bispectrum of COBE-DMR Four Year Maps}  
\author{H\aa vard B. Sandvik and  Jo\~{a}o
Magueijo\\
\affiliation{Theoretical Physics,  
Imperial College, Prince Consort Road, London SW7 2BZ, UK}}
%\affiliation{h.sandvik@ic.ac.uk}}

%\affil{Theoretical Physics,  
%Imperial College, Prince Consort Road, London SW7 2BZ, UK} 
\maketitle
 
\begin{abstract} 
We extend a previous bispectrum analysis of the COBE-DMR 
4 year maps,  allowing for the presence of
correlations between all possible angular scales. We find that 
the non-Gaussian signal found earlier for bispectrum components
including adjacent modes does not extend to triplets of modes with 
larger separations. Indeed  for all separations $\Delta\ell >1$ we find 
that the COBE-DMR data is very Gaussian. The implication seems to be 
that the previously detected non-Gaussian scale-scale correlation falls 
off very quickly with mode separation.
\end{abstract} 
 
%\keywords{Cosmology: cosmic microwave 
%background -- theory -- observations} 
 
\section{Introduction} 
\label{introduction} 

The issue of whether or not the primordial fluctuations are Gaussian
is of fundamental importance  in theories of structure formation.
Gaussian large-scale primordial fluctuations 
are believed to be the hallmark of the simplest inflationary scenarios.  
In these theories vacuum quantum fluctuation, with Gaussian statistics,
are stretched by a period of exponential expansion to give rise to the 
large scale fluctuations we observe today. In contrast, 
topological defect theories, 
which are also possible candidates for structure formation
\cite{string1,string2}, are believed 
to produce non-Gaussian density fields.  Non-Gaussianity is also 
believed to be associated with some non-minimal inflationary models
\cite{Salopek:1992jq,Linde:1997gt,Contaldi:1999jr}, such as those
generating isocurvature  
fluctuations \cite{peebles}. 

Several recent \cite{fmg,pvl98,magueijo1}
papers have shown indications of non-Gaussianity in the
cosmic microwave background (CMB) temperature fluctuations,
as measured by the COBE-DMR experiment. The first of these detections\\
\cite{fmg} was later found to be due to an experimental 
systematic \cite{banday}, and the second \cite{pvl98} to an error of method 
\cite{belen}. The third claim \cite{magueijo1}, however, remains unassailable.
In that work a bispectrum analysis was carried out, using adjacent
scales $\{\ell-1,\ell,\ell+1\}$ as components of the bispectrum.
A significant deviation from Gaussianity was found, with a high
confidence level. 

An obvious question is whether this non-Gaussianity
extends to components correlating scales with a larger separation,
say $\{\ell-\Delta\ell,\ell,\ell+\Delta\ell\}$, with $\Delta\ell >1$.
The purpose of this paper is to answer this question. 
As we shall see the non-Gaussian signal found in
\cite{magueijo1} does not appear to extend to larger separations
$\Delta\ell$.  

\section{Method}
In this letter we examine the  possibility of deviations from 
Gaussianity in terms of the bispectrum, which should be
zero for a Gaussian process. Let us expand 
the temperature fluctuations $\frac{\Delta
T}{T}({\bf n})$  in Spherical Harmonic functions:
\begin{eqnarray} 
\frac{\Delta T}{T}({\bf n})=\sum_{\ell m}a_{\ell m}Y_{\ell m}({\bf n}) 
\label{almdef} 
\end{eqnarray} 
A Gaussian probability distribution function is completely described by
the first two moments, $\langle a_{\ell m} \rangle$ and $\langle
a_{\ell m}a_{\ell'm'} \rangle $ i.e. all higher moments can be obtained from
the mean value and the variance.  This does not hold for Non-Gaussian
functions.  The Bispectrum, a rotationally invariant cubic form
associated with the third moment, $\langle
a_{\ell m}a_{\ell'm'}a_{\ell'' m''} \rangle$, is therefore a possible
test of Gaussianity.  It is given by
%The bispectrum is related the 3-point correlation
%function is therefore a common test for Non-Gaussianity.  
%The coefficients $a_{\ell m}$ can be 
%The coefficients $a_{\ell m}$ may then be combined into  
%rotationally invariant multilinear forms (see \cite{santa} 
%for a possible algorithm). The most general 
%cubic invariant is the bispectrum,  and is given by 
\begin{eqnarray} 
{\hat B}_{\ell_1\ell_2\ell_3}&=&\alpha_{\ell_1\ell_2\ell_3} 
\sum_{m_1m_2m_3}\left  
( \begin{array}{ccc} \ell_1 & \ell_2 & \ell_3 \\ m_1 & m_2 & m_3 
\end{array} \right ) a_{\ell_1 m_1}a_{\ell_2 m_2} a_{\ell_3 m_3} 
\nonumber \\ 
\alpha_{\ell_1\ell_2\ell_3}&=&\frac{1}{(2\ell_1+1)^{\frac{1}{2}} 
(2\ell_2+1)^{\frac{1}{2}}(2\ell_3+1)^{\frac{1}{2}}}\left ( 
\begin{array}{ccc} \ell_1 
 & \ell_2 & \ell_3 \\ 0 & 0 & 0 
\end{array} \right )^{-1} 
\label{bispec} 
\end{eqnarray} 
where the $(\ldots)$ is the Wigner $3J$ symbol.  
In \cite{magueijo1} correlations between multipoles separated by $\Delta
\ell = 1$, $\{\ell-1\, \ell \,\ell+1\}$  were investigated.  In this letter we
complete this work by using the other possible separations, $\Delta l
= 2,3,4,5, \text{ and } 6$, which fit within the signal dominated range
of $\ell$ probed by the COBE-DMR experiment.
Selection rules require $\ell_1 + \ell_2 +
\ell_3$ to be even, $|\ell_i - \ell_j| \le \ell_k \le \ell_i+\ell_j$
and  $m_1+m_2+m_3=0$.  From these constraints we see that a
suitable chain of correlators is $\hat A_\ell^{\Delta
\ell}=B_{\ell-\Delta \ell\, \ell \,\ell+\Delta \ell}$, with $\ell$
even.

We then follow precisely the same procedure documented in 
\cite{magueijo1} and \cite{fmg}, apart from considering the more 
general ratio
\begin{eqnarray} 
J^{ \Delta \ell}_\ell &=& { {\hat A}_{\ell}^{\Delta \ell} 
\over ({\hat C}_{\ell-\Delta \ell})^{1/2}({\hat C}_{\ell})^{1/2} 
({\hat C}_{\ell+\Delta \ell})^{1/2}} 
 \label{defI} 
\end{eqnarray} 
where ${\hat C}_\ell=\frac{1}{2\ell+1}\sum_m|a_{\ell m}|^2$.  
This quantity is dimensionless, and is invariant under rotations 
and parity. The same data is examined; the inverse
noise-variance-weighted average maps of the 53A,53B,90A and 90B
\emph{COBE}-DMR channels, with monopole and dipole removed, at
resolution 6, in galactic and ecliptic pixelization. We use the   
extended galactic cut of \cite{banday97}, and  
\cite{benn96} to remove most of the emission from the 
plane of the Galaxy. 
To  estimate the $J^{ \Delta \ell}_\ell$s we set 
the value of the pixels within the galactic cut to 0 and  
the monopole and dipole of the cut map to zero.  
We then integrate the 
map multiplied with spherical harmonics  
to obtain the estimates of the 
$a_{\ell m}$s and apply equations \ref{bispec} and \ref{defI}. 

The observed $J^{ \Delta \ell}_\ell$s are to be compared with their
distributions $P(J^{ \Delta \ell}_\ell)$ as inferred from Monte
Carlo simulations in which  
Gaussian maps are subject to DMR noise and galactic cut.
In simulating DMR noise we take into account the full
noise covariance matrix, as described in \cite{corn}. This includes
correlations between pixels $60^\circ$ degrees apart.

\section{Results and conclusions}

\begin{figure} 
\begin{center}
\psfig{file=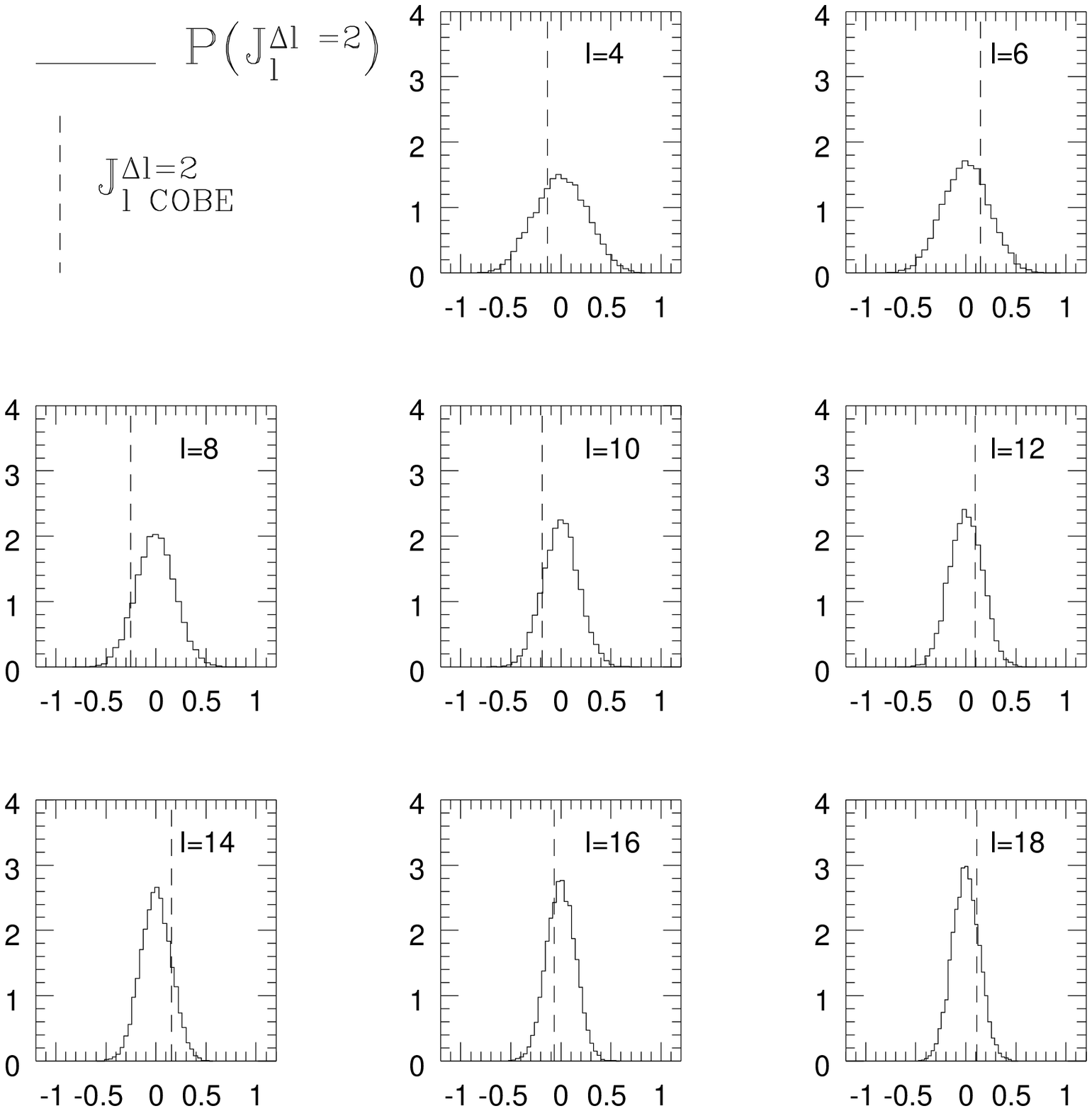,width=12cm}
\end{center}
\caption{The vertical thick dashed lines represent the values  
of the observed $J^{ \Delta \ell = 2}_\ell$, for different values of
$\ell$. 
%s for different values
%of the inter $\ell$ separation, $\Delta \ell=1,2,..,6$.  
The solid
lines are the corresponding probability distribution functions of the
$J^{ \Delta \ell=4}_\ell$s for a Gaussian sky with extended galactic
cut and DMR noise, as inferred from 10000 realizations.}
\label{fig1} 
%\end{figure}
\end{figure}

The comparison between $P(J^{ \Delta \ell}_\ell)$ and the 
observed $J^{ \Delta \ell}_\ell$ for various $\Delta \ell$
is displayed in figs. ~\ref{fig1} - ~\ref{fig4}.  The observed values
agree well 
with the probability distributions from the Monte Carlo
distributions: implying that as far as $\Delta\ell>1$ 
bispectrum components are concerned the data supports Gaussianity.

To quantify just how well the observations agree with Gaussianity, we do a
goodness of fit statistic on the data.  We calculate the $\chi^2$
value for the different $\ell$ separations, and compare with the
respective distributions $F(\chi^2)$. The rationale behind considering
separate $\chi^2$ was explained in \cite{magueijo1}\footnote{This is to be
contrasted with more dubious philosophy towards combining separate
tests expressed in \cite{teg,kog}.}, and for the distributions
 $P(^{ \Delta \ell}_\ell)$ found (which are very close to Gaussians)
it makes sense to use the standard $\chi^2$:
\begin{equation}
\chi^2 = \frac{1}{N}{\sum_{i=1}^N} \frac{\left( x_i-\mu_i\right)}{\sigma_i^2}
\end{equation}
where $N$ is the number of different observables and  $\mu_i$ and
$\sigma_i^2$ are the
expectation value and the variance of the $i$th observable respectively. 
This reduces to the expression proposed in \cite{fmg} in our case.
The $F(\chi^2)$ are then calculated through 
further Monte Carlo simulations, so that possible correlations
between the different $\ell$ may be taken into account.  The $\chi^2$
distributions and corresponding observed values are shown in fig
~\ref{fig5}. 

As we can see $\chi^2\approx 1$ for all separations $\Delta\ell >1$,
with deviations well within the spread expected from $F(\chi^2)$.
This is to be contrasted with the results obtained for $\Delta\ell=1$,
that is bispectrum components measuring correlations between adjacent
modes. For these $\chi^2\ll 1$, meaning that the observed $J_\ell$
do not display the scatter around zero expected from a Gaussian 
distribution. We now found that this distinctive non-Gaussian signal
does not extend to bispectrum components with  $\Delta\ell>1$.
The implication seems to be that the inter-mode correlations
suggested by the findings of \cite{magueijo1} fall off quickly
with the scale separation, something which conforms to theoretical
prejudice. 

In a recent paper \cite{kog} a variety of components
of the bispectrum was studied in the context of 
a semi-analytical texture model.
Such a model is at the very least oversimplified, and probably
has little to do with real-life textures. In any case that work 
stresses the importance of analysing off-diagonal components,
as we have done in this paper.

%The results are displayed in Fig.~\ref{fig1}.  Apart from the results
%for $\Delta \ell =1$, which are identical to the results in
%\cite{magueijo1}, the results seem to be Gaussian, nicely scattered
%around the expectation value.
\begin{figure} 
\begin{center}
\psfig{file=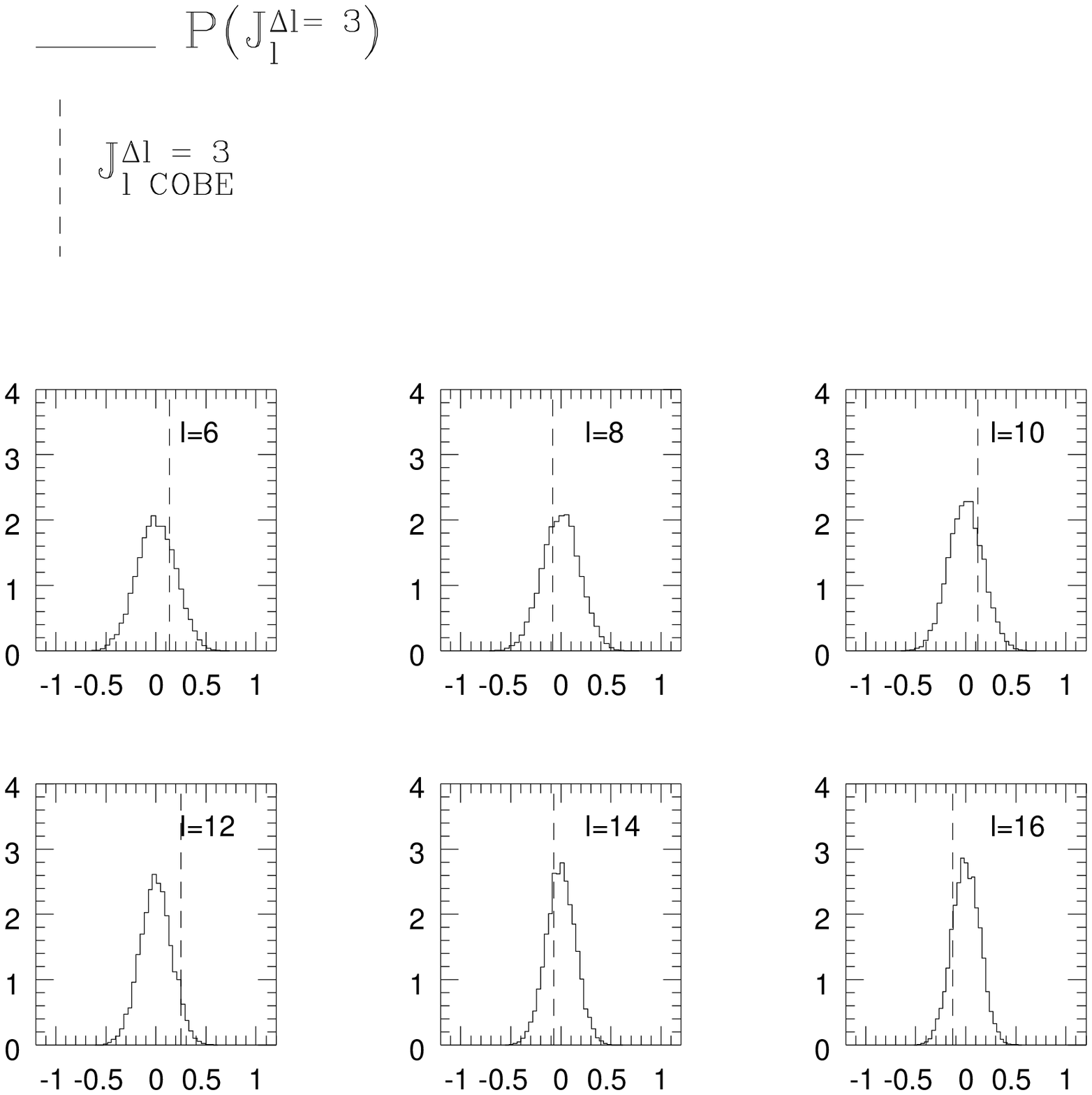,width=12cm}
\end{center}
\caption{The vertical thick dashed lines represent the values  
of the observed $J^{ \Delta \ell = 3}_\ell$, for different values of
$\ell$. 
%s for different values
%of the inter $\ell$ separation, $\Delta \ell=1,2,..,6$.  
The solid
lines are the corresponding probability distribution functions of the
$J^{ \Delta \ell=4}_\ell$s for a Gaussian sky with extended galactic
cut and DMR noise, as inferred from 10000 realizations.}
%\caption{The vertical thick dashed lines represent the values  
%of the observed $J^{ \Delta \ell}_\ell$s for different values
%of the inter $\ell$ separation, $\Delta \ell=1,2,..,6$.  The solid
%lines are the corresponding probability distribution functions of the
%$J^{ \Delta \ell}_\ell$s for a Gaussian sky with extended galactic
%cut and DMR noise, as inferred from 10000 realizations.}
%\label{fig5} 
%\end{figure}
\end{figure}

\begin{figure} 
\begin{center}
\psfig{file=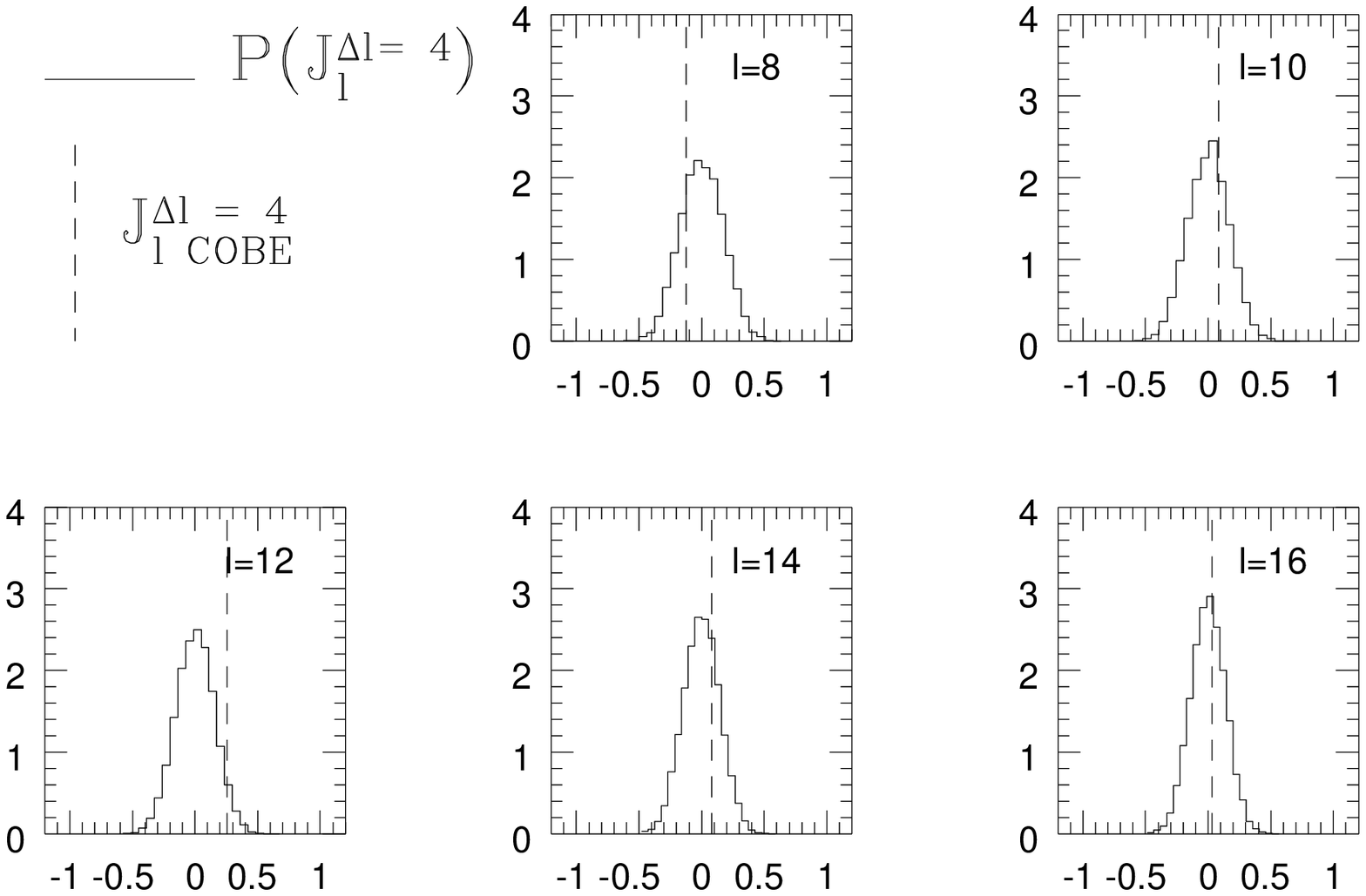,width=12cm}
\end{center}
\caption{The vertical thick dashed lines represent the values  
of the observed $J^{ \Delta \ell = 4}_\ell$, for different values of
$\ell$. 
%s for different values
%of the inter $\ell$ separation, $\Delta \ell=1,2,..,6$.  
The solid
lines are the corresponding probability distribution functions of the
$J^{ \Delta \ell=4}_\ell$s for a Gaussian sky with extended galactic
cut and DMR noise, as inferred from 10000 realizations.}
\end{figure}

\begin{figure}
\begin{center}
\begin{tabular}{c}
\psfig{file=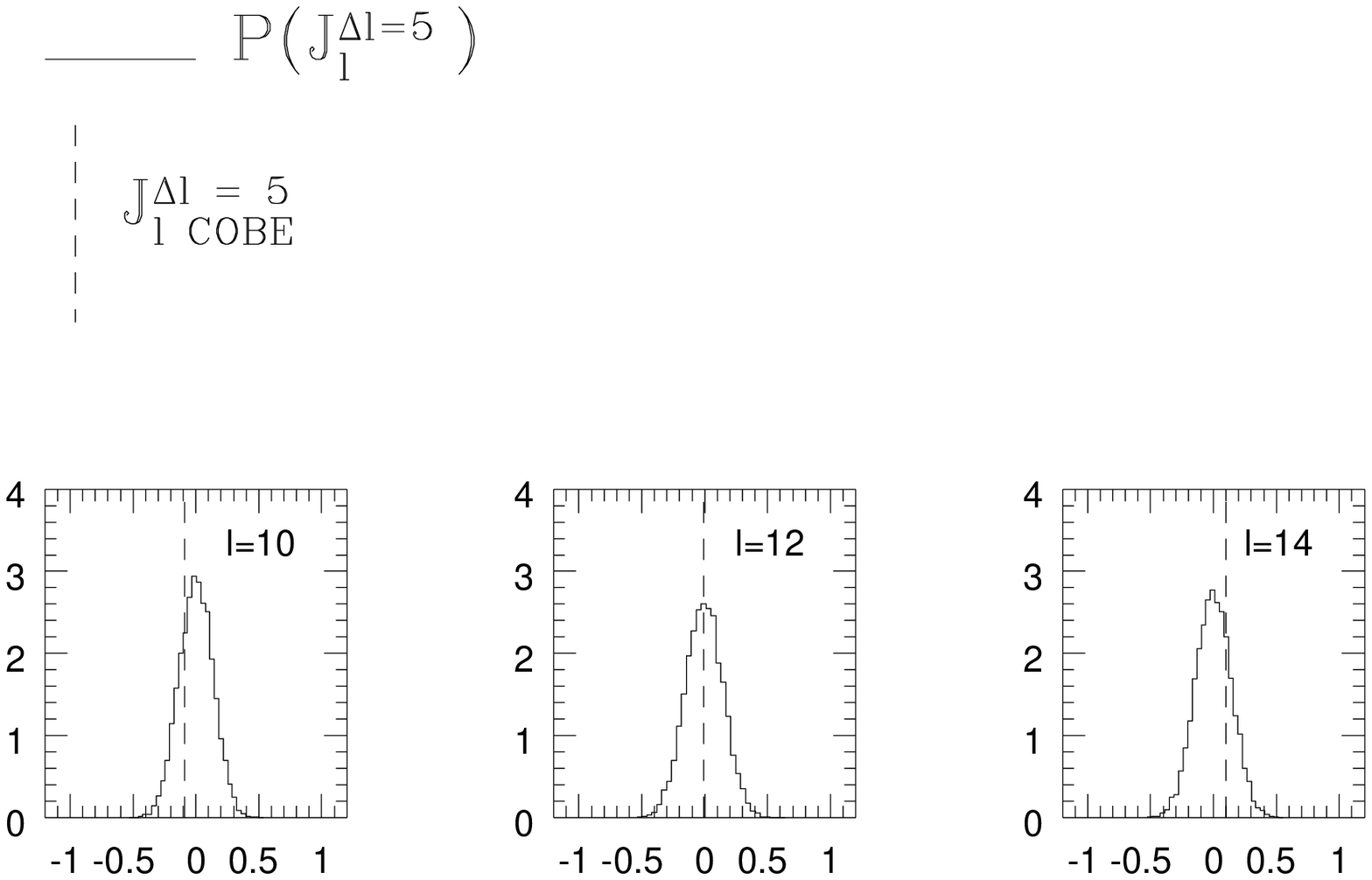,width=8cm}
%\caption{The vertical thick dashed lines represent the values  
%of the observed $J^{ \Delta \ell}_\ell$s.  The solid lines are the
%probability distribution functions of $J^{ \Delta \ell}_\ell$ for a
%Gaussian sky with extended galactic cut and DMR noise, as inferred
%from 25000 realizations. The plots are for different values
%of the inter $\ell$ separation, $\Delta \ell=1,2,..,6$.} 
%\label{fig5} 
%\end{figure}
%&
\\
\psfig{file=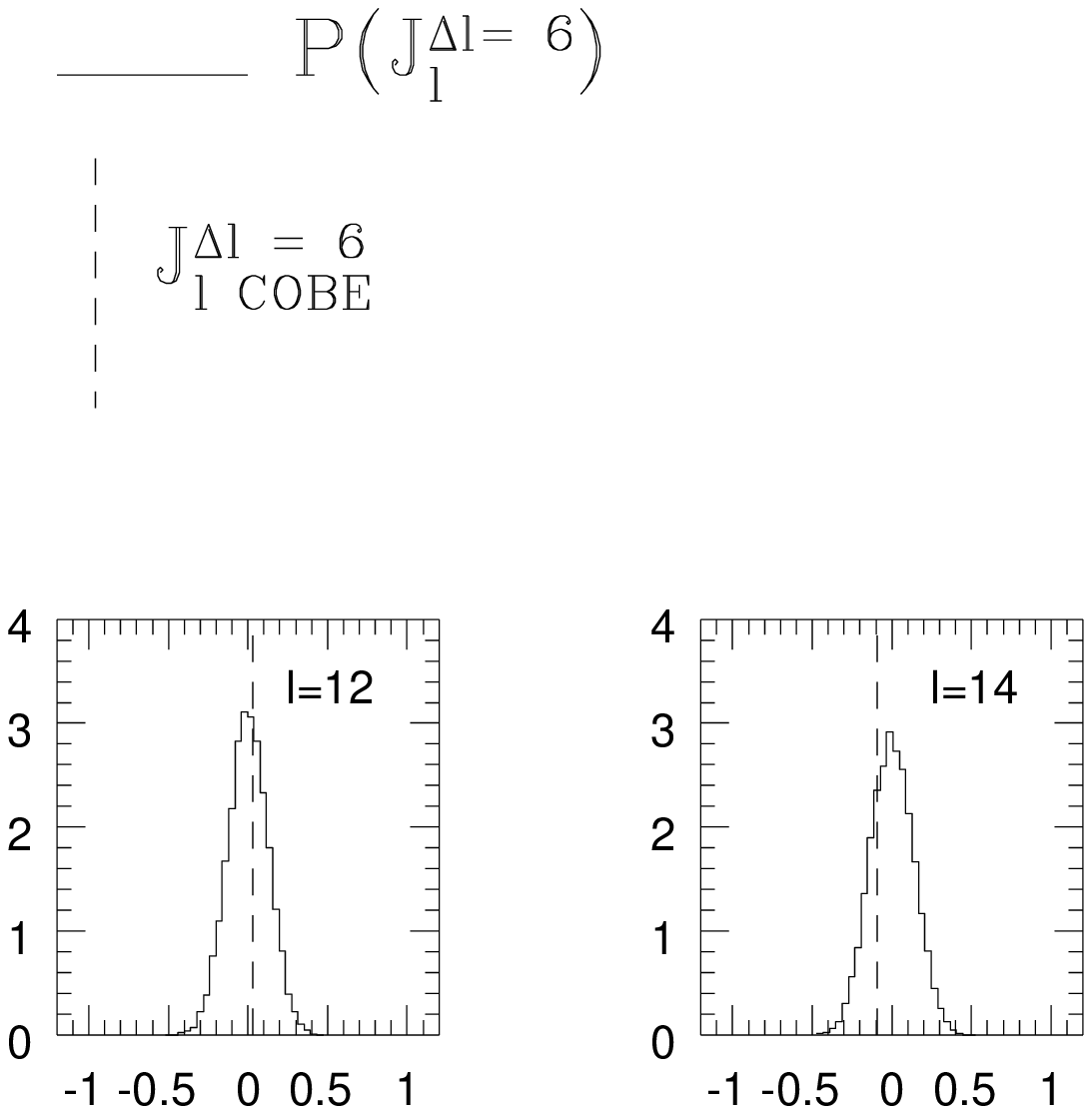,width=8cm}
\end{tabular}
\end{center}
\caption{The vertical thick dashed lines represent the values  
of the observed $J^{ \Delta \ell}_\ell$s, $\Delta \ell = 5$, and $6$
for different values 
of $\ell$.  The solid
lines are the corresponding probability distribution functions of the
$J^{ \Delta \ell}_\ell$s for a Gaussian sky with extended galactic
cut and DMR noise, as inferred from 10000 realizations.}
\label{fig4} 
%\end{figure}
\end{figure}

\begin{figure}
\begin{center}
\begin{tabular}{cc}
\psfig{file=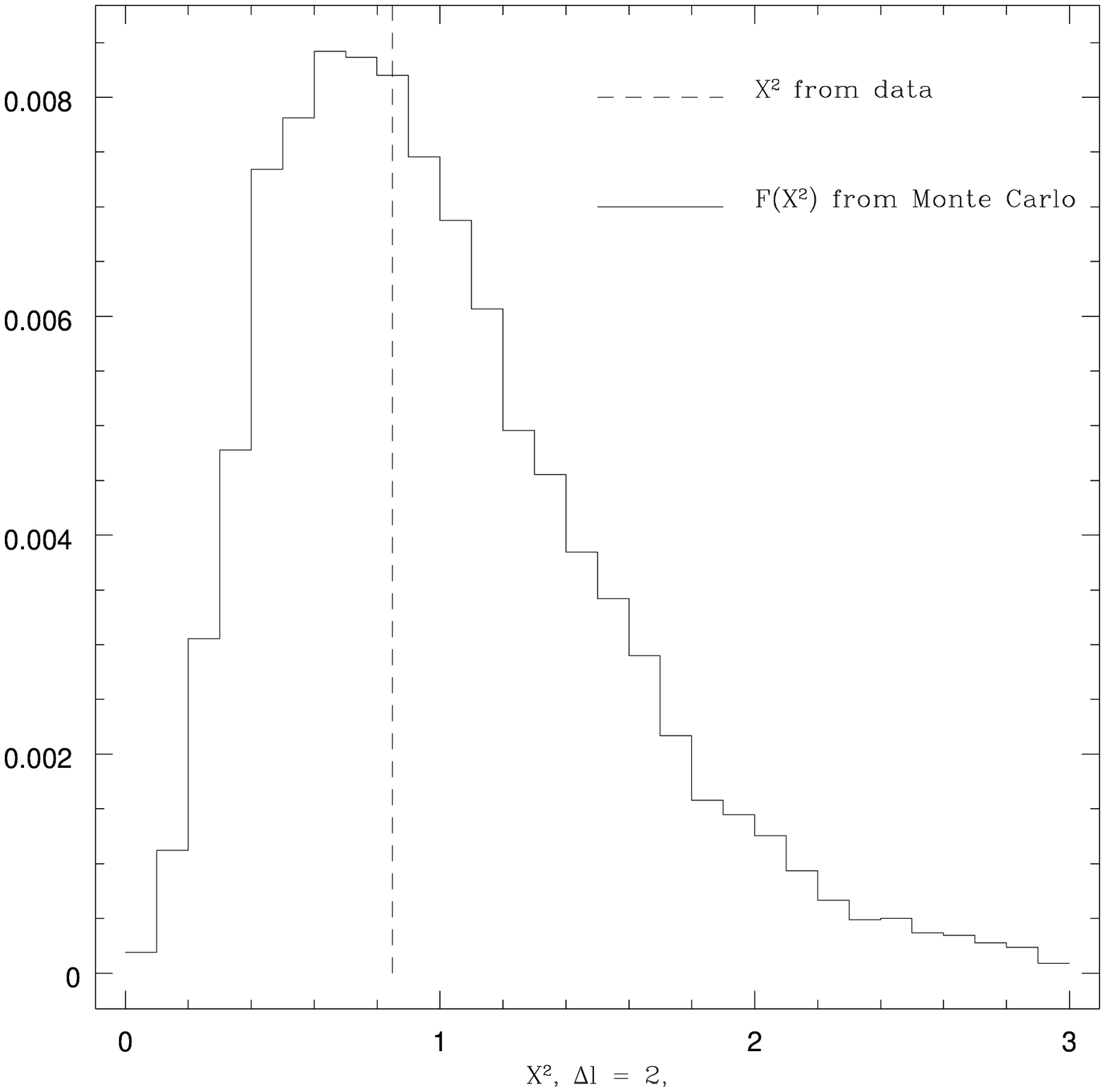,width=6cm}
&
\psfig{file=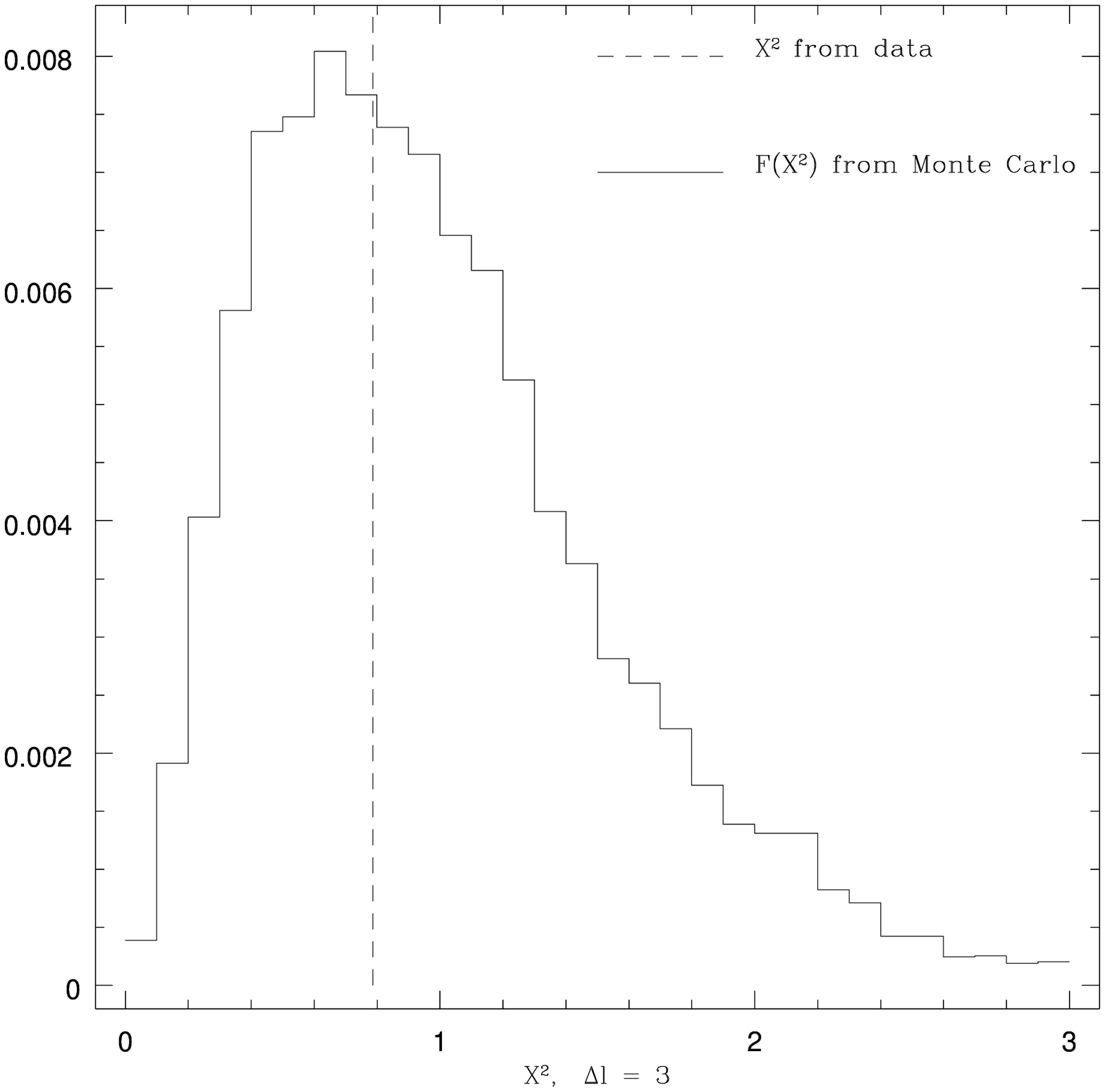,width=6cm}
\\
\psfig{file=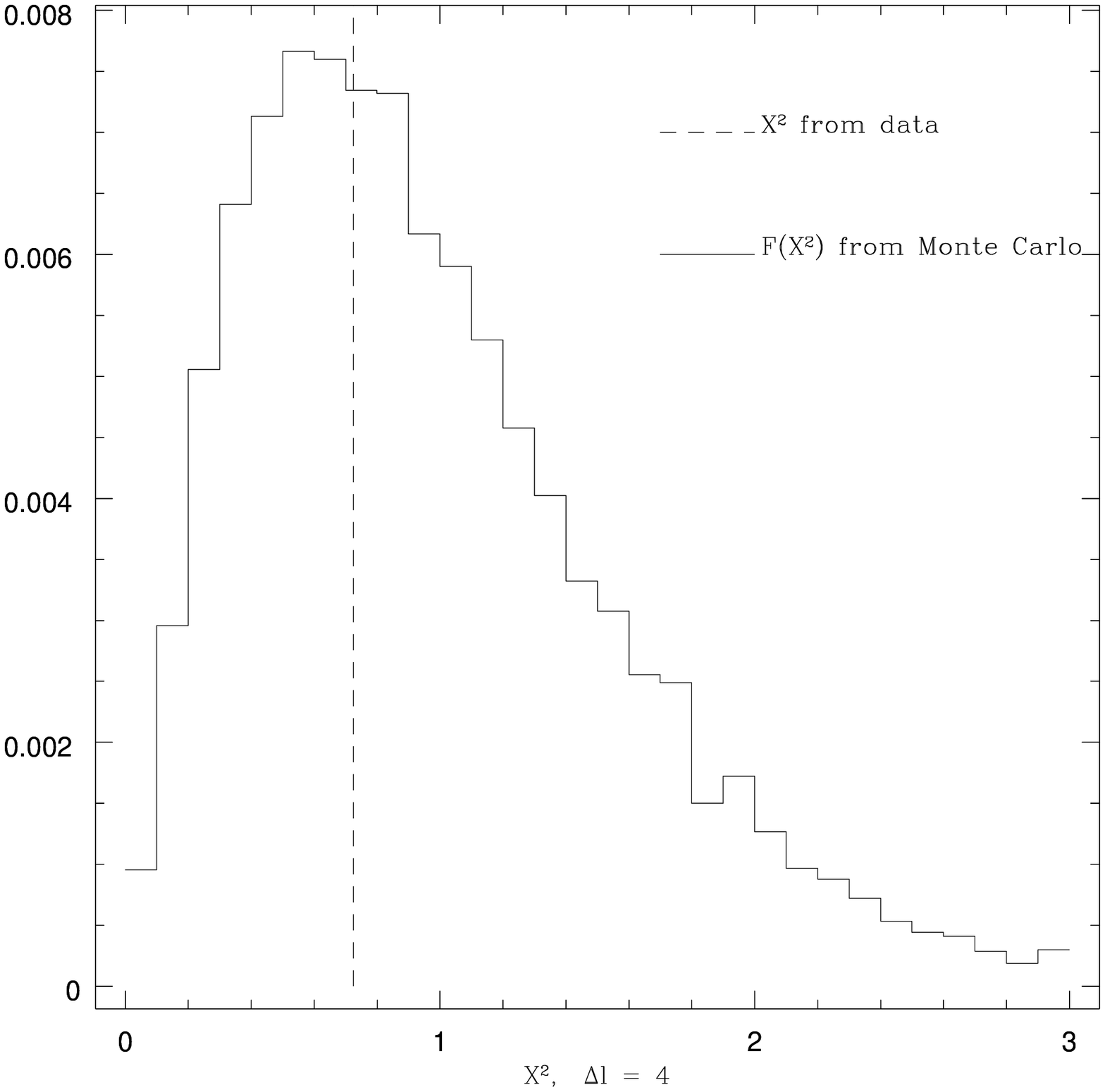,width=6cm}
&
\psfig{file=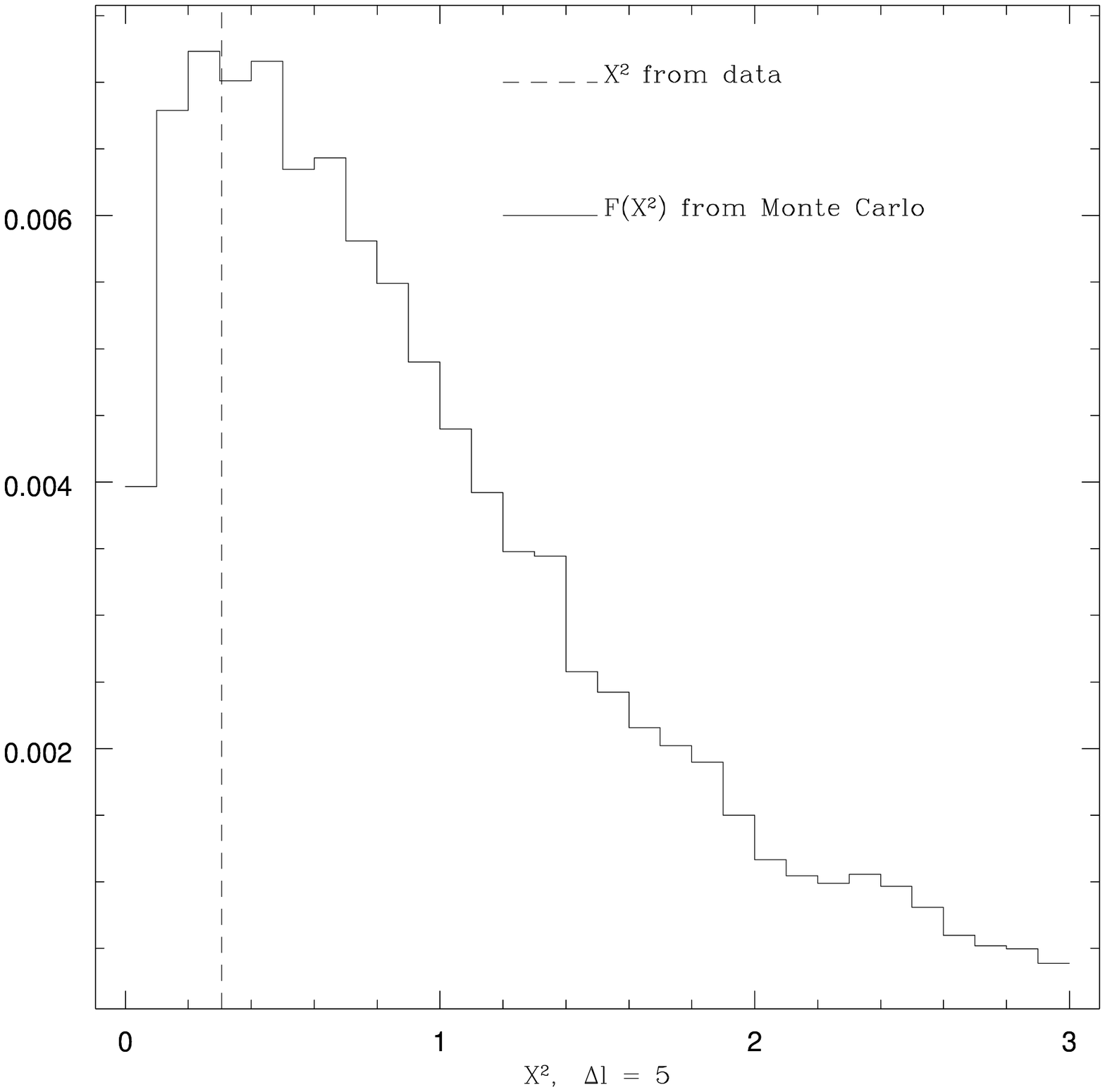,width=6cm}
\\

\end{tabular}
\end{center}
\caption{The dashed lines are observed values of $\chi^2$ from the
COBE data for different $\ell$ separations $\Delta \ell = 2,3,4$ and
$5$.  The solid lines are 
corresponding probability distribution functions for $\chi^2$ for a
Gaussian sky with extended galactic cut and DMR noise, as inferred
from 10000 realisations.}
\label{fig5}
\end{figure}

\section*{ACKNOWLEDGEMENTS} 
H.B. Sandvik would like to thank The Research Council of Norway for
financial support. This work was performed 
on COSMOS, the Origin 2000 supercomputer owned by the UK-CCC and 
supported by HEFCE and PPARC.

\end{document}